\documentclass[a4paper,alpha-refs]{aejstyles}

\journal{aej}

\usepackage{graphicx}
\usepackage{siunitx}

\newcommand{\eqn}[1]{Equation~(\ref{#1})}   
\newcommand{\fig}[1]{Figure~\ref{#1}}      
\newcommand{\sex}{\texttt{SourceExtractor}} 

\newcommand\degree       {{\ifmmode^\circ\else$^\circ$\fi}}  
\newcommand\arcm         {{\ifmmode {'\ }\else$'     $\fi} } 
\newcommand\arcs         {{\ifmmode{''\ }\else$''    $\fi} } 

\newcommand\arcspt       {{$\buildrel{\prime\prime}\over .$}}
\newcommand{\new}          {\textbf}  
\newcommand\n            {\noindent}

\newcommand\AHaH         {{\it AHaH}}

\newcommand\cge          {{$_ >\atop{^\sim}$}}
\newcommand\cle          {{$_ <\atop{^\sim}$}}

\def\eg{{\it e.g.}}

\newcommand{\del}[1]{\relax}%
\newcommand{\DEL}[1]{\relax}%
\newcommand{\DELETED}[1]{\relax}%
{\relax}%

\title{Interactive Cosmology Visualization Using the Hubble UltraDeep Field Data 
in the Classroom}

\author[1,\authfn{1}]{Liam J. Nolan}
\author[1]{Mira R. Mechtley}
\author[1]{Rogier A. Windhorst}
\author[1]{Karen Knierman}
\author[1]{Teresa A. Ashcraft}
\author[1]{Seth  H. Cohen}
\author[1]{Scott Tompkins}
\author[2]{Lisa M. Will}

\affil[1]{School of Earth and Space Exploration, Arizona State University, Tempe, AZ 85287-1404}
\affil[2]{San Diego City College, San Diego, CA 92101}

\authnote{\authfn{1}ljnolan@asu.edu}

\papercat{Research Article}

\runningauthor{Nolan et al.}

\jvolume{00}
\jnumber{0}
\jyear{2017}

\begin{document}

\begin{frontmatter}
\maketitle
\begin{abstract}
We have developed a Java\footnote{\url{http://java.sun.com}}-based teaching tool,
``Appreciating Hubble at Hyper-speed'' (``\AHaH''), intended for use by students
and instructors in beginning astronomy and cosmology courses, which we have made
available online\footnote{\url{http://ahah.asu.edu/download.html}}. This tool lets the
user hypothetically traverse the Hubble Ultra Deep Field (HUDF) in three
dimensions at over $\sim500\!\times\!10^{12}$ times the speed of light, from
redshifts $z\!=\!0$ today to $z\!=\!6$, about 1 Gyr after the Big Bang. Users may
also view the Universe in various cosmology configurations and two different
geometry modes -- standard geometry that includes expansion of the Universe, and
a static pseudo-Euclidean geometry for comparison. In this paper we detail the
mathematical formulae underlying the functions of this Java application, and
provide justification for the use of these particular formulae. These include the
manner in which the angular sizes of objects are calculated in various cosmologies,
as well as how the application's coordinate system is defined in relativistically
expanding cosmologies. We also briefly discuss the methods used to select and
prepare the images in the application, the data used to measure the redshifts of
the galaxies, and the qualitative implications of the visualization -- that is,
what exactly users see when they ``move'' the virtual telescope through the
simulation. Finally, we conduct a study of the effectiveness in this teaching
tool in the classroom, the results of which show the efficacy of the tool, with over 
$\sim$90\% approval by students, and provide justification for its further use 
in a classroom setting.
\end{abstract}


\end{frontmatter}


\section{Introduction} \label{intro}

In beginning astronomy courses, many non-science majors appear to have a
significant lack of understanding -- even after taking the introductory courses
-- of basic concepts such as wavelength, the electromagnetic spectrum, the
speed of light, lookback time, redshift, and the expansion of the Universe. We
believe this lack of concept acquisition or retention represents a significant
shortcoming of the currently available teaching tools. While pictures, figures,
and other static media are certainly effective at communicating many concepts,
they tend to be poor at showing these effects in three dimensions, or those
that evolve over time (\eg\ \citealt*{Sadaghiani2011}). Since virtually all cosmological 
effects require very large time or distance scales to become apparent, a different 
teaching medium is needed in this case. This is reflected by the well-established need in astronomy education to support the development of spatial skills, which strongly correlates with performance in all STEM disciplines \citep{Cole2018}.

In addition, the education landscape is changing at a breakneck pace around us,
with online learning quickly becoming a preferred option for reasons of
convenience, access, and affordability, especially to students of limited
means. At its extreme, in the case of large-scale threats to safety, online
learning becomes mandatory, as has been so dramatically underlined by the
recent outbreak of COVID-19. As millions of students around the world moved
from learning in a classroom to learning at home, it became evidently clear
that modern classes require new tools for education. Nowhere is this more
needed than in the laboratory-type complement to lecture-based learning, and
online virtual tools stand to serve this purpose exceptionally well (\eg\ \citealt*{Hoeling2012}).

``Appreciating Hubble at Hyper-speed'' (\AHaH) is an educational tool that aims
to address these issues of concept acquisition and retention by providing a
visual and interactive learning medium. The project uses data from the Hubble Space Telescope (HST)
Cycle 12 Project ``GRAPES'' \citep[Grism-ACS (Advanced Camera for Surveys) Program for 
Extragalactic Science;][]{Pirzkal2004} to build a redshift-sorted database of over 5000
galaxies within the Hubble Ultra Deep Field (HUDF). As a simplified, brief explanation, if one begins Hubble's Law of the Universal Expansion (as per the FLRW model\footnote{For further discussion of the differences between the Hubble's redshift-distance and velocity-distance laws, see \citet{Harrison1993}.}),

\begin{equation}\label{Dhubble}
 D = \frac{v}{H_0}
\end{equation}

\n and combine with the Doppler approximation as valid for small redshifts $v\!\approx\!cz$, one obtains

\begin{equation}\label{D}
 D = \frac{v}{H_0} \approx \frac{cz}{H_0}
\end{equation}

\n and can see a galaxy's distance is proportional to redshift, $z$, divided by Hubble's constant, $H_0$\footnote{We use $H_0\approx68$ km/s/Mpc throughout \citep{Planck2018}.}. As these galaxies range from
redshift $z\!\approx\!0.05$ to $z\!\approx\!6$, their distances span nearly 90\% of the
history of the Universe \citep{YanWindhorst2004, Bouwens2006, Cohen2006,
Windhorst2011}. Since these data represent the deepest optical images of the
Universe ever obtained, they are thus uniquely suited to help students understand
the effects of the expanding Universe. 

\begin{figure}[bt!]
\centering
\includegraphics[width=\linewidth]{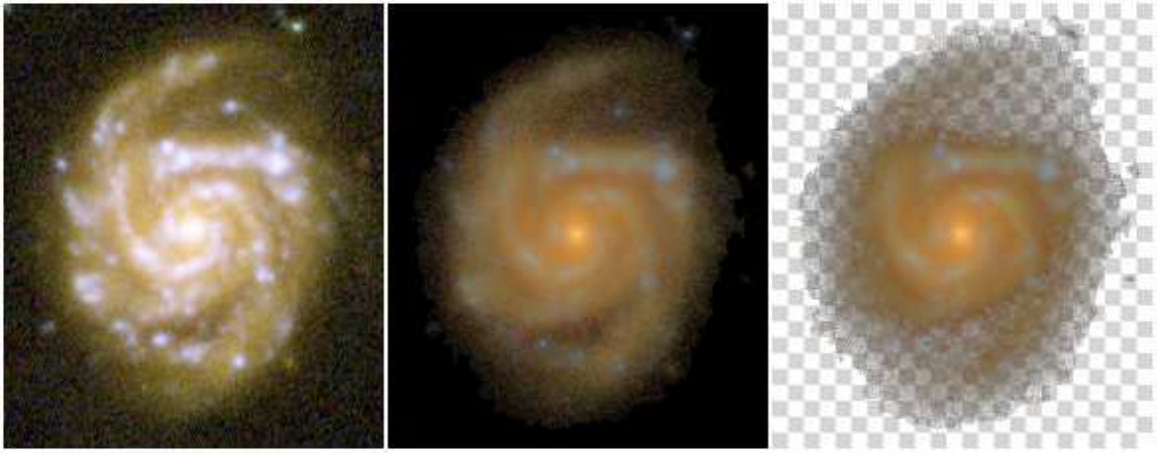}
\caption{A comparison of three images of HUDF galaxy 7556. The left image is
that from the original STScI release, clearly showing the bright, burned-out
knots characteristic of the standard logarithmic image stretch. The center
image is our prepared image using the arcsinh stretch described by
\citet{Lupton2004}, as it appears in the \AHaH\ application. The right image is
our prepared image against an artificially imposed chessboard pattern, showing
the included transparency. Note that pixels outside the source are all
completely transparent, since they have been removed entirely using the \sex\
segmentation map.}
 \label{figcomparison}
\end{figure}

\begin{figure}[bt!]
\centering
\includegraphics[width=\linewidth]{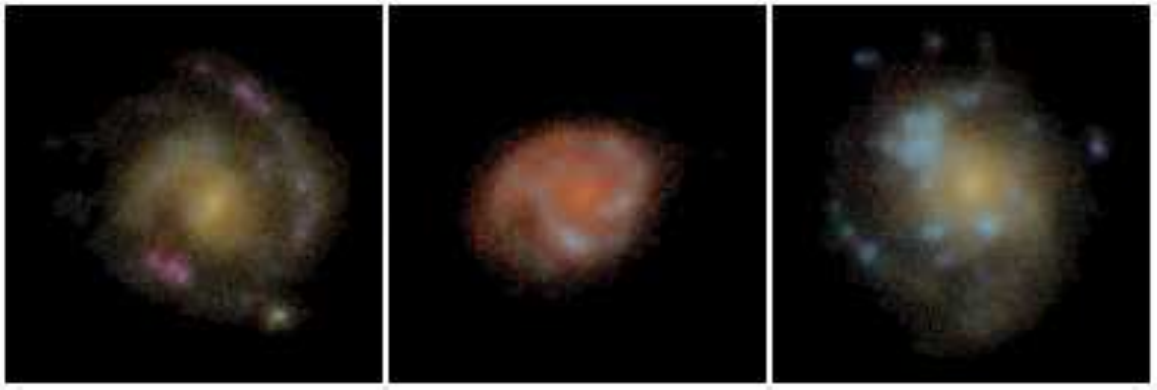}
\caption{Our prepared images of three galaxies from the HUDF, using the
arcsinh stretch described by \citet{Lupton2004}. Shown are galaxy 3180 (left),
galaxy 5805 (center), and galaxy 6974 (right).}
 \label{figthreegalaxies}
\end{figure}

It is worth noting that for the \AHaH\ Application, we developed a custom-balanced RGB version of the original HUDF image to eliminate bright areas being "burned out" and lacking fine detail \citep{Lupton2004}.  This custom version is then displayed as semi-transparent, and images are attached to photometric redshifts measured using HyperZ \citep{Bolzonella2000} using photometry from the publicly available HST optical and near-infrared images \citep{Thompson2005}, as well as spectro-photometric redshifts from \citet{Ryan2007}. A comparison of the original STScI color images and our prepared images is shown in \fig{figcomparison}, and one can see examples of galaxies processed in this way in \fig{figthreegalaxies}.  We explain this in more detail in \S~\ref{prep} (Appendix B).

\del{\section{Background}
\label{sec:background}

The background section should be written in a way that is accessible to researchers without specialist knowledge in that area and must clearly state---and, if helpful, illustrate---the background to the research and its aims. The section should end with a brief statement of what is being reported in the article.}

\section{Theoretical Formalism}
\label{formulae}

As discussed in detail by \citet{Wright2006}, there are a number of different
methods for calculating distances in cosmology. For our purposes, the most
meaningful of these is the comoving radial distance, $D_R$, representing the
spatial separation of an object and an observer with zero peculiar velocity at a
common cosmic time. This distance takes into account the expansion of the
Universe, and so is more useful when dealing with distances on very large scales
(and thus very large look-back times), as is the case with most galaxies in the
HUDF. Henceforth, we shall adopt the convention of referring to the comoving
radial distance from Earth to a galaxy as $D_R$ in Gpc\footnote{\ \ 1 Gpc =
10$^9$ pc, and 1 pc = 3.26 lightyear.}, and the comoving coordinate distance
between two arbitrary points in the coordinate system as $r_{ij}$.

As we begin to define different cosmological distances, we wish to note that there are many different distance definitions and naming conventions in cosmology which ultimately come down to three distances. For discussions and examples of the variety used in the literature see, for example, \citet{Kristian1966}, \citet{Weinberg1972}, \citet{Peebles1993}, \citet{Longair1995}, \citet{Hogg1999},\new{, \citet{Ellis2007}, \citet{Etherington2007}, and \citet{Ellis2009}}. We adopt the names comoving radial distance ($D_R$), angular size distance ($D_A$), and luminosity distance ($D_L$), \new{with $D_A$ and $D_L$ being} related by the ''distance duality relation."  Regardless of terminology or convenient approximations (i.e. \eqn{D}), we carefully use the correct distance formalism ``under the hood" to produce the visualization in \AHaH .

Returning to the development of the tool, we also wish to calculate the angular sizes of objects as they would be observed
from redshifts other than zero. To do so, we need a formula for the angular size
distance, $D_A$. That is, the distance which satisfies the equation $d\!\approx\!\theta
D_A$ for an object with transverse linear diameter $d$ subtending an angle $\theta$ in
the field of view at any redshift in any relativistic cosmology. Following \citet{Ribeiro2005}, we note that the correct relation is between solid angle and $l^2$ over $D_A^2$, but in the small area of the HUDF we may linearize as previously stated.  This is also important to allow the tool to run at a reasonable speed on consumer computers. In a simple
Euclidean space, this is the same as the radial distance, but again we must take
into account the expansion (and possible curvature) of the Universe, so we must
use a separate equation for $D_A$ in the \AHaH\ tool.  Details on these definitions and equations are given in \S~\ref{derivations} (Appendix A).

In addition, we need to consider how we wish to define the coordinate system for
the objects within the Java tool. Although we have very deep HST imaging data
that allow us to show how the Universe has changed over time, all of these data
were collected at a common time (2003/2004--2014). Moreover, the principal
distance measure that we have available, the comoving radial distance $D_R$, also
assumes a common cosmic time. Thus the most sensible coordinate system is one
with three spatial dimensions that makes all calculations for a common cosmic
time, viz. when the data were collected. We can then contract the distances in
this ``comoving coordinate system'' as necessary to simulate observations from
redshifts greater than zero. The question remains of how we should derive such
coordinates from the data that we have in such a way that they will be useful to
us -- this is discussed in \S~\ref{coordinates} in Appendix A, prior to deriving the
equations.  We detail our calculations in full in \S~\ref{derivations} 
(Appendix A), which we include for both completeness and instructional purposes 
-- as these calculations should be comprehensible in an intermediate 
undergraduate-level cosmology course -- but are not necessary to follow for 
a demonstration of the tool's educational utility.  Ultimately, we are able 
to develop a relationship between angular size distance and redshift, as 
well as a coordinate system, which is logical with our available data.  We 
use these relationships (with a few simplifications for computational 
efficiency) to simulate the ``motion" of our \AHaH\ camera.

\section{Standard Display Mode}
\label{std disp}

\begin{figure}[bt!]
\centering
\includegraphics[width=\linewidth]{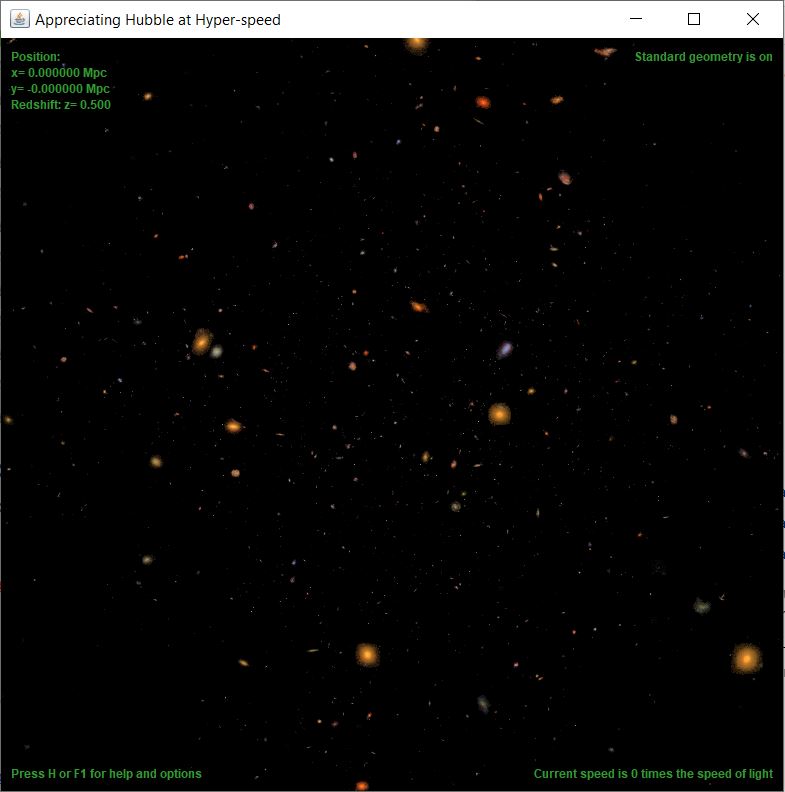}
\caption{The HUDF data as viewed from redshift $z=0.5$ in the \AHaH\
application, using standard geometry mode, which properly calculates angular
sizes. Note how the image is dominated by luminous red early-type galaxies at
moderate redshifts of z\cle 1, where the Universe is older than 6 billion years.}
 \label{figahahrs0.5}
\end{figure}

While some might argue that the equations in \S~\ref{derivations} 
(Appendix A) speak for themselves, we
believe it is very instructive to consider the qualitative implications to their user 
-- that is, a description of what exactly we see when we ``move'' the
camera in the Java application. For the sake of completeness, we will also
detail a number of cosmological effects that have been omitted from the
application due to technical limitations. An example of the standard display
mode is shown in \fig{figahahrs0.5}.

When we move the camera to a certain position in the HUDF (X, Y, Z) data cube, we are in
general viewing the Universe as it would appear from that point and at that
redshift. We must qualify this statement by noting that the simulation accounts
\textit{only} for cosmological effects of changing the camera position -- no other
dynamical, gravitational lensing, evolutionary, or other cosmological or physical effects are
simulated. In this sense, \AHaH\ thus truly, though hypothetically, allows the
user to travel through the Universe at ``hyper-speed.'' The highest virtual
speed of $\sim$500$\times$10$^{12}$ times the speed of light that AHaH uses to
zoom into the HUDF database allows the observer to travel from z=0 to z\cle 6
in a fraction of a minute, rather than in the $\sim$12.9 Gyr needed if the
maximum ``travel'' speed were really limited to the speed of light $c$, as in the real Universe. 

As is covered in many introductory physics and astronomy classes, we use the Euclidean small angle approximation (SAA), where for a fixed object length $l$ and a small angle $\theta$:

\begin{equation}\label{theta}
 \theta = tan{(\frac{l}{D})} \approx \frac{l}{D},
\end{equation}

\n where the distance to the object is $D$.  In relativity, we use essentially the same equation, but replace $D$ with $D_A$, where $D_A$ is a complex function of redshift with a maximum at $z\approx1.65$.  Thus the relativistic SAA is:

\begin{equation}\label{theta_approx}
 \theta \approx \frac{l}{D_A}.
\end{equation}

\begin{figure}[bt!]
\centering
\includegraphics[width=\linewidth]{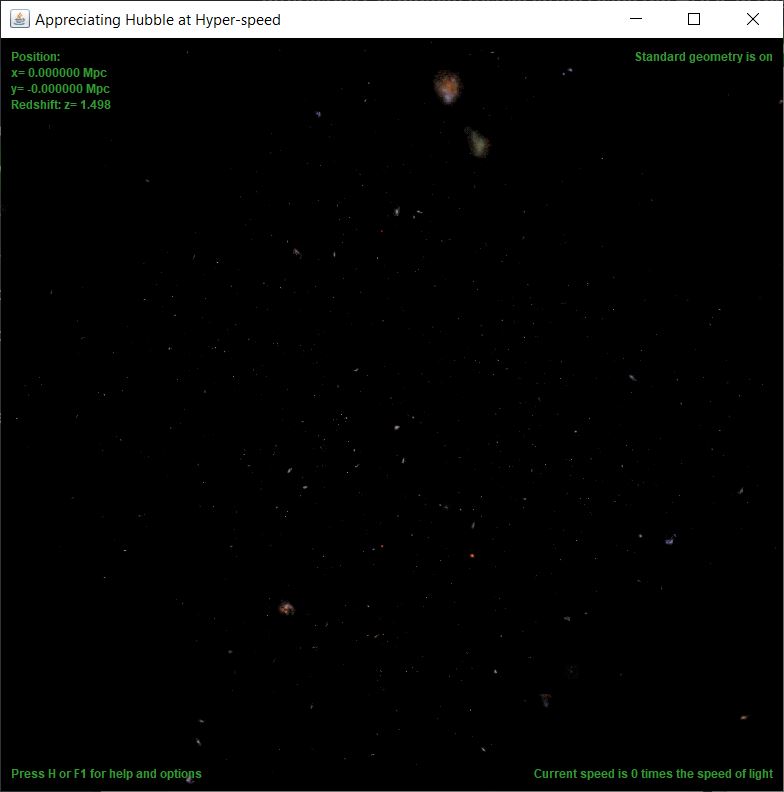}
\caption{The HUDF data as viewed from redshift $z=1.5$ in the \AHaH\
application, using standard geometry mode. Note how this image is dominated by
blue irregular and merging star-forming galaxies, and that all red ellipticals
of \fig{figahahrs0.5} are now ``behind us''. This Universe at z\cge 1.5 is
the actively star-forming Universe, where the Universe is younger than 4 billion years.}
 \label{figahahrs1.5}
\end{figure}

The somewhat counter intuitive relationship (called the $\Theta$ -- z relation) between an object's angular size and
its redshift in Relativistic Cosmology is readily apparent in the standard display mode. 
If a user slowly increases the redshift of the camera, high redshift objects will begin to
decrease in angular size and move toward the center of the display, eventually
reaching a minimum angular size (at redshift z$\simeq$1.65 in standard
$\Lambda$CDM cosmology \citep{Planck2018}), and then increasing. Also visible are the effects of
galaxy evolution and merging over time. For example, when viewing the Universe
from redshift $z\!=\!0.5$ as in \fig{figahahrs0.5}, there are many large spiral
and elliptical galaxies visible. Note how the Universe is dominated by luminous
red, early-type galaxies at moderate redshifts of z\cle 1 in \fig{figahahrs0.5}.
However, when viewing the Universe from redshift $z\!=\!1.5$ as in
\fig{figahahrs1.5}, the \AHaH\ screen is dominated by small and compact blue
galaxies. When zooming into the data at z\cge 1.5, many of these objects are blue
irregular, merging and/or star-forming galaxies. In \fig{figahahrs1.5}, all red
ellipticals of \fig{figahahrs0.5} are now ``behind us''. This Universe at z\cge
1.5 is indeed the actively star-forming Universe, i.e., the first 4 billion years after the Big Bang.

It should be noted that the application does not make calculations for
cosmological surface brightness dimming or changes in color due to redshift or
spectral evolution. While certainly feasible to simulate, performing such image
manipulation techniques on large numbers of galaxies in real-time is currently
too difficult for consumer computers. Moreover, we must also recall that the HUDF
data are limited in both magnitude and effective horizon by what could be
observed from low Earth orbit. When we view the data from redshifts other than
zero, we would expect to see more galaxies overall -- including fainter galaxies
-- than are present in the current HUDF data. We could choose to simulate these
objects as extensions of our data set if desired, but we felt this would not
be particularly instructive, and could lead to potential confusion, as fainter
galaxies would have to be continuously simulated in increasing numbers by the
computer below Hubble's detection limit when traveling from z=0 today to z\cle 6
($\sim$1 Gyr after the Big Bang). Moreover, such simulations have a high degree
of uncertainty, and by significantly increasing the size of the data set, would
add prohibitively to the computation times. Likewise, we have chosen not to
simulate galaxies outside of the original field, which would of course enter the
camera's field of view as the user pans around.

\section{Static Geometry Mode}
\label{static geom}

\begin{figure}[bt!]
\centering
\includegraphics[width=\linewidth]{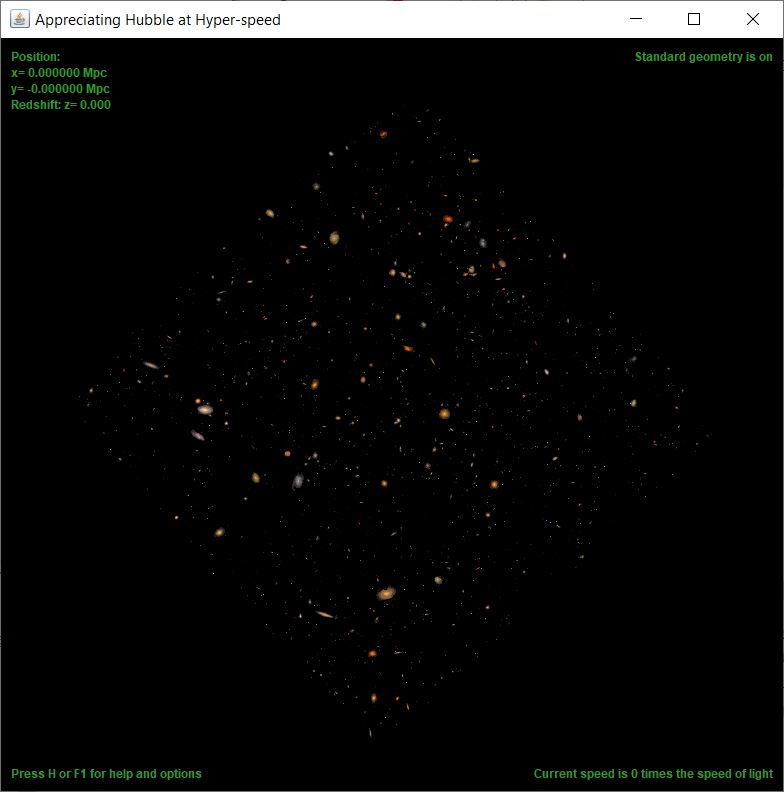}
\caption{The HUDF data as viewed from redshift $z=0$ in the \AHaH\
application, using standard geometry mode. This displays the entire HUDF, for
comparison with Figures \ref{figahahEuclid} and \ref{figahahLambda}.}
 \label{figahahrs0}
\end{figure}

\begin{figure}[bt!]
\centering
\includegraphics[width=\linewidth]{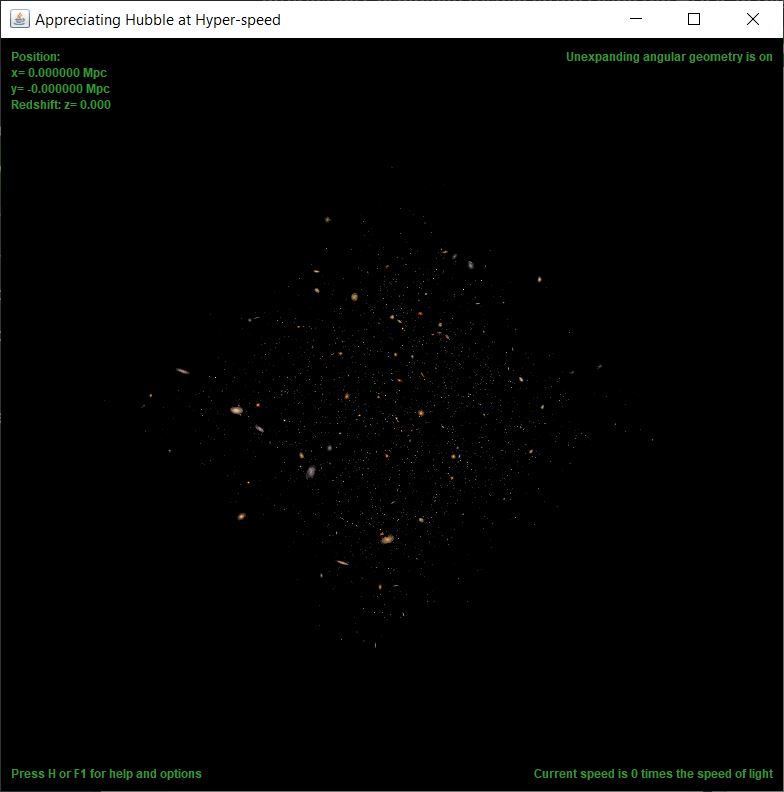}
\caption{The HUDF data as viewed from redshift $z=0$ in the \AHaH\ application,
using static geometry mode. This displays the entire HUDF, and one can see the
evident ``contraction" of the field of galaxies due to the lack of non-Euclidean
geometry in the real HUDF data. (That is, the HUDF data in a  relativistically
expanding cosmology -- when shown in Euclidean geometry -- compresses all
objects towards the image center, since the square cylindrical volume is now not
undergoing the expansion, as it should).}
 \label{figahahEuclid}
\end{figure}

When a user presses the ``G'' key in the \AHaH\ Java tool, they are told that they are
viewing the simulation with ``unexpanding angular geometry'' turned on. What
this means specifically is that angular sizes as derived above are no longer
affected by the scale factor or curvature of the Universe. After we develop our
original coordinates, as in \eqn{X Definition}, all calculations for angles are
simply done with $\theta\!=\!\theta_E$. This has the visual effect of all
galaxies appearing smaller and closer to the center of the viewport (as can be seen in \fig{figahahEuclid} as compared to \ref{figahahrs0}, since all
initial angles have been contracted by an expansion factor of $(1+z)$ 
(when the curvature energy density $\Omega_K$ is
zero; see \S~\ref{extreme} for an explanation of the main energy densities at play in 
Relativistic Cosmology). In this static case, galaxies will also simply increase in angular size as
we approach them, as opposed to the angular sizes of high-redshift objects in the
real $\Lambda$CDM Universe \citep{Planck2018}, which decrease, reach a minimum, and then {\it increase}
again in angular size as the camera's redshift {\it increases}.

This static mode of viewing the simulation has no physical analogue -- it is
simply meant to convey to the user that there are non-Euclidean aspects of the
Universe's geometry, and that the angular sizes that we observe in the present have
been made larger due to the universal expansion. One should note that this
display mode only considers expansion as it relates to angular size -- the
comoving radial distance is still calculated using the redshift and curvature
factors that would not be present in a strictly Euclidean Universe. That is, in
the static display mode, we assume that the Hubble Law distance,
$D\!=\!v/H_0\!\approx\!(c/H_0)z$, is simply a Euclidean distance unrelated to
expansion. This is primarily because our method of calculating the comoving
radial distance relies using all object redshifts, which is a phenomenon 
specific only to an expanding Universe, and is therefore the only way we could
calculate the distances for all galaxies in other -- hypothetical -- Universes.

\section{Exploration of Extreme Cosmologies}
\label{extreme}

\begin{figure}[bt!]
\centering
\includegraphics[width=\linewidth]{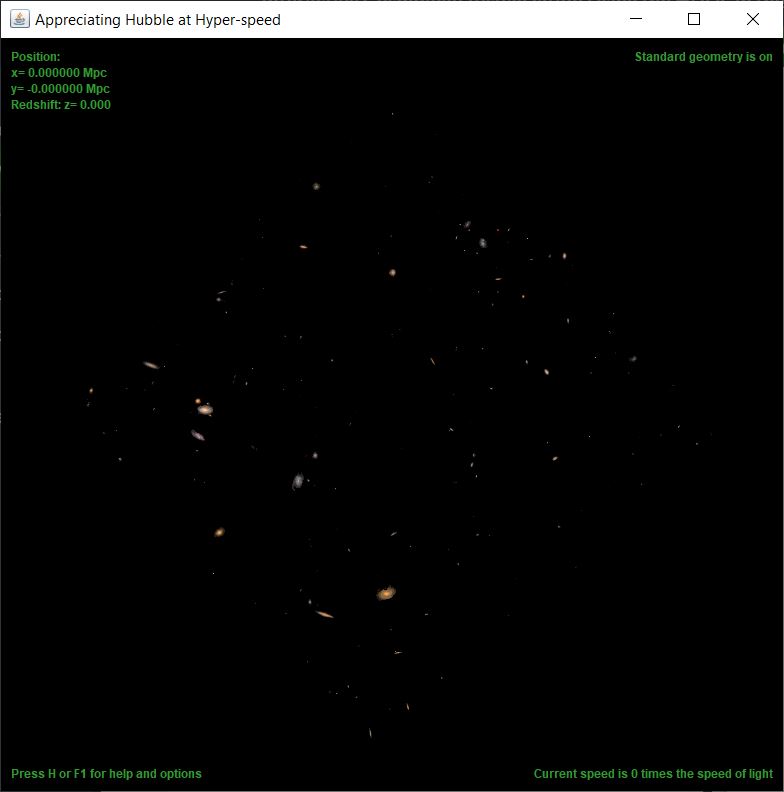}
\caption{The HUDF data as viewed from redshift $z=0$ in the \AHaH\
application, using standard geometry mode, with a vacuum energy density parameter $\Omega_{\Lambda}$ value
increased from the default value of $0.763$ to $2.0$. One can see that the
extreme value causes most galaxies that are visible in the standard
$\Lambda$CDM cosmology \citep{Planck2018} to disappear due to vastly increased
expansion in this case. This value was selected because noticeably higher
values of $\Omega_{\Lambda}$ cause all galaxies to disappear entirely -- we observers likely live in a relativistic Universe with a ``little Lambda."}
 \label{figahahLambda}
\end{figure}

One additional capacity of the \AHaH\ software we wish to note is the representation
of wildly different universes from our own, in terms of the cosmological
parameters used in the calculations above. As we have already noted, many
students have difficulty with relatively esoteric concepts of the energy
density parameters of the Universe, and how different portions dominate cosmic
behavior over time.  To briefly summarize for the reader, $\Omega_{M}$, 
$\Omega_{R}$,  $\Omega_{\Lambda}$, and $\Omega_{K}$ are the main cosmological parameters, which are 
the fractions  of the Universe's total average energy density that are attributable 
to matter (M), radiation (R), dark energy ($\Lambda$), and the curvature of the 
spatial geometry (K), respectively. It is assumed these are the only relevant 
contributions to the total energy density $\Omega_{Tot}$, i.e. $\Omega_{Tot}$ = 
$\Omega_M + \Omega_\Lambda + \Omega_R + \Omega_K$ (with a spatially flat Universe 
having  $\Omega_{Tot} \equiv 1$ with $\Omega_K$ = 0). However, instead of leaving 
this to verbal and written descriptions (which can be 
obtuse), relying on a student's ability to mentally visualize these
concepts, \AHaH\ allows the student to form an approximate image of any
cosmology with whatever parameters one could desire. By varying a single
parameter, one can see the dramatic (or sometimes less so) impact on our vision
of the Universe. We present as an example a massive increase in
the vacuum energy density (or Einstein's cosmological constant) $\Omega_\Lambda$ 
from 0.763 to 2.0 (as in \fig{figahahLambda}), which of course
results in a Universe rapidly pushed outwards until just a few galaxies 
hypothetically remain in Hubble's field-of-view. This value was selected
because at noticeably higher values of $\Omega_\Lambda$, {\it all} galaxies
disappear nearly completely from view sooner. Of course, just as with the static 
viewing mode described above, this change in parameters does not produce a perfect 
representation  of a Universe with these parameters, and what is simulated by the 
software has no physical analogue. However, the changes in display under extreme 
values of different constants is still instructive as to the effects of these 
different constants on the structure of the Universe.

\section{Integration in the Classroom}
\label{classroom}

In order to test the effectiveness of this software in a real education
environment, we conducted a study in astronomy classes at Arizona State
University in the northern hemisphere fall semester of 2020. It should be noted that due to the outbreak of COVID-19, ASU classes 
were moved into a partially online Synchronous format for part/all of the northern hemisphere fall 2020 semester, thus introducing an additional 
variable as compared to a typical semester. This pandemic underlines the necessity 
of developing online laboratory tools which complement virtual instruction, thus we 
felt continuing with our study was particularly appropriate.

\subsection{Methods}
\label{methods}

We selected the introductory astronomy labs for non-majors (AST 113 northern hemisphere Fall 2020) and for astronomy majors (SES 123 northern hemisphere Fall 2020) for our
study in order to focus on the target population mentioned above -- students
first being introduced to this type of astronomy content. Both classes are typically composed of first-year students, so the level of preparation is similar, though students pursuing astronomy majors tend to come into college with more background information in our experience. We modified the
curriculum of all sections of the lab in two different orders, and compared our 
results between the two orderings of labs. We requested the ASU Institutional Review
Board (IRB) for approval to make this study in the AST 113 and SES 123 Lab classrooms, and their 
approval, containing all conditions for the study, was filed with ASU. Students who 
enrolled in these modified lab sections were informed that the new programs with our
virtual tool would be available, and that this would not affect the course 
expectations or standards.

We selected several exercises in the typical class curriculum in both labs for replacement
with \AHaH\ materials based on the advice of the course instructor as to which
exercises could have their essential materials folded into other activities or
the class lecture component. We then introduced in replacement of these exercises
a series of exercises designed by members of our team for \AHaH, all of which are
available online\footnote{\url{http://ahah.asu.edu/exercises.html}}.
Students then conducted these exercises either on AST 113/SES 123 Lab or their own home
computers, with the same resources available to them as other activities in class
(i.e. technical and subject-matter support from Teaching Assistants, instructor, etc.).
Reasonable accommodations for learning and access difficulties as pursuant to
University policy were made for, \eg, those students who needed to take this
course remotely via the Zoom teleconference platform given their presence elsewhere during the COVID-19
pandemic.  In addition, for visually impaired students, we made available the 
Astronomy Sound of the Month webpage for January 2018
\footnote{\url{https://astrosom.com/Jan2018.php}}, which has a full display of 
the HUDF image and plays a tone when one moves the mouse over a given galaxy, of 
varying pitch depending on the distance to that galaxy. All lab sections 
were led for these activities by a combination
of Lab Teaching Assistants and the class instructor, as appropriate.

In some cases, material covered by these \AHaH\ exercises is not part of the
subject matter typically evaluated by the course. However, the course instructor 
believed there was sufficient relation to course content, and students were made 
aware of this slight alteration. These subjects included, \eg:
galaxy morphology, Hubble's Law, galaxy evolution, the $\Theta-z$ relation, and
cosmological parameters.

\subsection{Educational Materials}
\label{edumat}

The first of the two student activities (referred to as ``labs") we led students through in this class was on the subject of galaxy morphology\footnote{\url{http://ahah.asu.edu/exercises/galmorph.pdf}}.  Students use the \AHaH\ tool to ``fly" through the universe and inspect the morphology of galaxies, either from an instructor-provided list or whichever they find interesting.  In the latter case, students are encouraged to pick a variety of galactic morphologies, as they are then asked to classify the galaxies by the Hubble Classification Scheme, and identify various other features.  Students then learn more about how different features can lead to conclusions about the populations of stars making up the galaxy.  The general learning outcomes produced are:

\begin{itemize}
    \item Identify galaxies by appearance
    \item Apply observed colors to the properties of its stars, and
    \item Understand the differences between the different Hubble classification types.
\end{itemize}

{The second lab was on the subject of ``Hubble's Law"\footnote{\url{http://ahah.asu.edu/exercises/hubble.pdf}}, by which we mean the linear relationship found by Hubble between distance and redshift at low redshifts.  In a similar fashion to the galaxy morphology lab, students move through the HUDF to inspect galaxies either from a list or of their choice, but in this case they use the built-in information pop-up on the galaxy to learn its redshift and comoving radial distance, $D_R$. Students plot their data points, and then use the ``Hubble Law" (the combination of the FLRW model Hubble Law and the Doppler approximation as given in \eqn{D}) to calculate $H_0$ from the slope of their graphs. A spreadsheet which would work out distances for higher redshifts was made available to students who wished to look at such galaxies.  The general learning outcomes produced are:}

\begin{itemize}
    \item Understanding the concept of redshift
    \item Obtaining and performing calculations with data to produce universal properties (such as $H_0$), and
    \item Gaining familiarity with the expansion of the universe.
\end{itemize}

\n These goals in particular address an established stumbling point in introductory undergraduate astronomy education - the curvature and behavior of the Universe - as described by \citet{Coble2018}.

The goals of these labs are well-described by the Anatomy of Disciplinary Discernment as laid out by \citet{Eriksson2014}.  For many students, the labs will serve to bring them from the first level, Disciplinary Identification (as students are expected to have been previously informed of terms such as ``galaxy" and ``redshift"), to the second level, Disciplinary Explanation, and be able to show how galactic properties distribute with redshift.  For other students the labs may be able to bring them to the third level, Disciplinary Appreciation, as they grasp the ``power" of Hubble's Law and other astronomical relations for astronomers to be able to describe the universe itself from the trends of galaxies.  We also see a few students actually take it upon themselves to learn more about how the standard calculations of the lab are approximations, and how these diverge at high redshifts - thus approaching the fourth and final level, Disciplinary Evaluation. We would also note that a third lab is available on our website concerning galaxy evolution\footnote{\url{http://ahah.asu.edu/exercises/galevol.pdf}}, but was not used in our study.

\subsection{Analysis and Results}
\label{ar}

\begin{figure}[bt!]
\centering
\includegraphics[width=\linewidth]{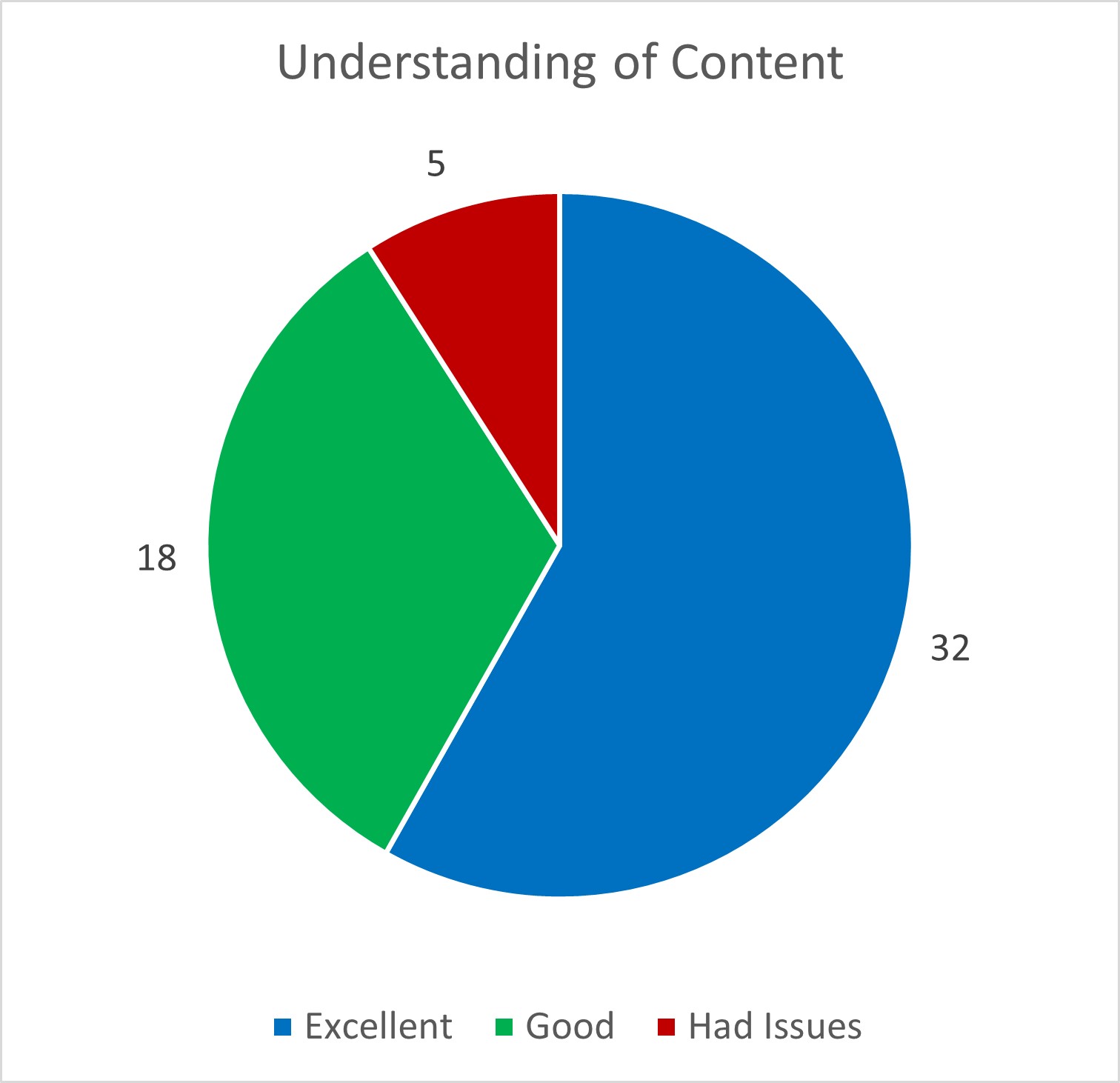}
\caption{Summary of results from northern hemisphere fall 2020 study, detailed in Table \ref{resultTable}, 
 specifically for question on how well the tool helped the students' understanding of the lab content.}
 \label{resultPie1}
\end{figure}

\begin{figure}[bt!]
\centering
\includegraphics[width=\linewidth]{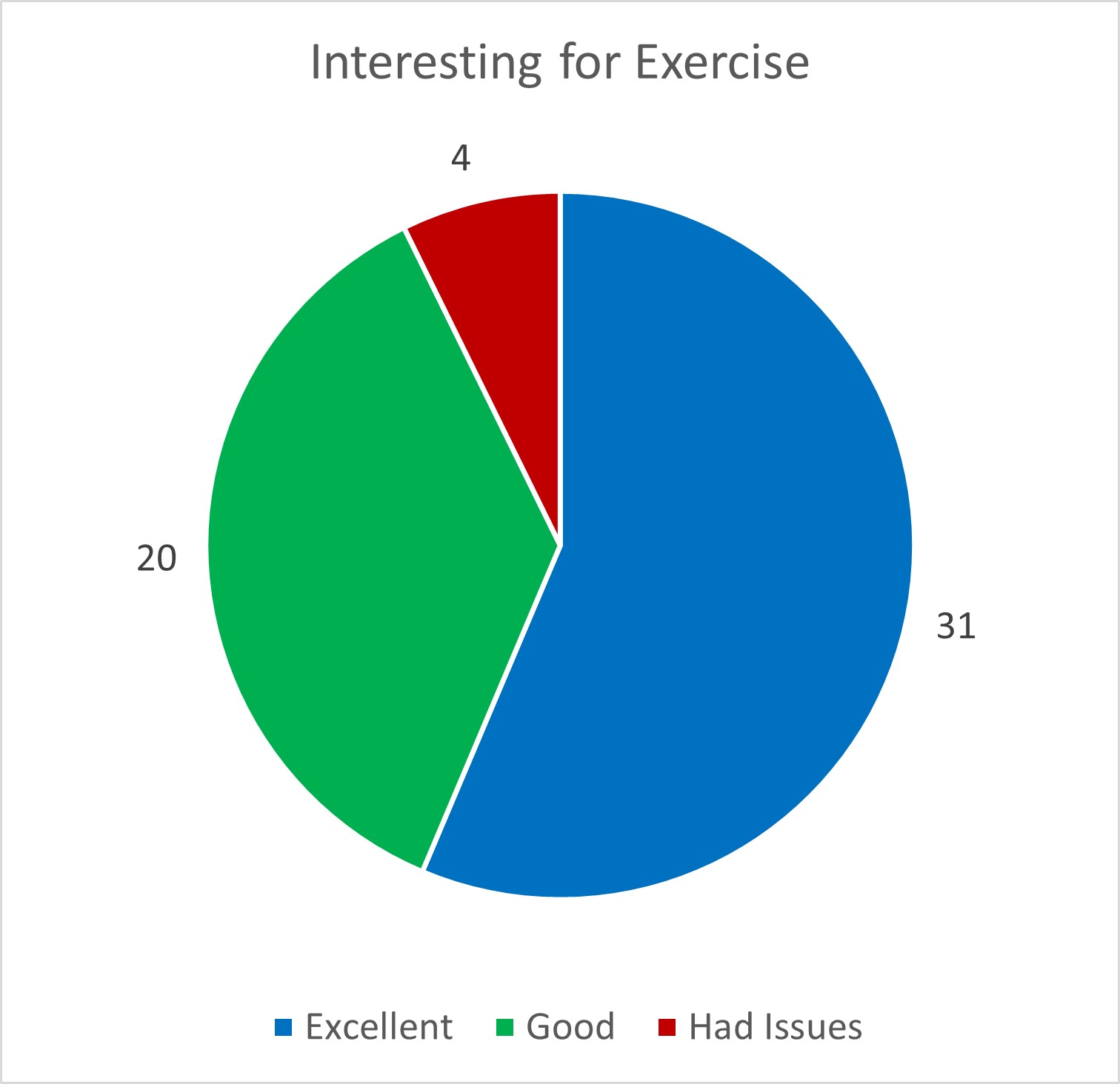}
\caption{Summary of results from northern hemisphere fall 2020 study, detailed in Table \ref{resultTable}, 
 specifically for question on how interesting students found the tool.}
 \label{resultPie2}
\end{figure}

\begin{figure}[bt!]
\centering
\includegraphics[width=\linewidth]{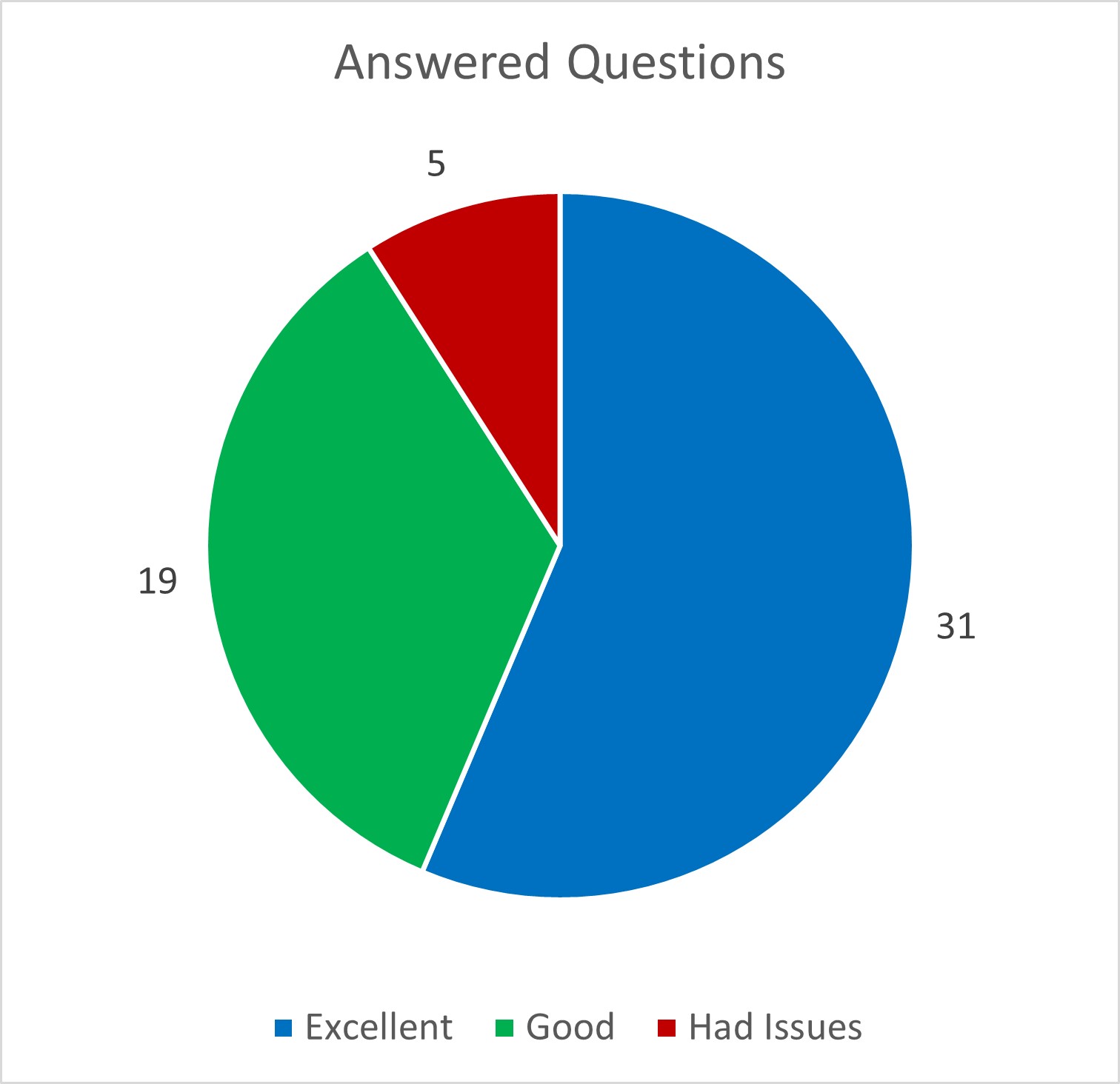}
\caption{Summary of results from northern hemisphere fall 2020 study, detailed in Table \ref{resultTable}, 
 specifically for question on how well the tool answered students' questions.}
 \label{resultPie3}
\end{figure}

At the end of the northern hemisphere fall 2020 semester class described above, we 
conducted a survey of students in the enhanced labs for their assessment of the
utility of the \AHaH\ tool, with questions shown in Table \ref{questionTable}, and 
complete results shown in Figures \ref{resultPie1}, \ref{resultPie2}, and \ref{resultPie3}, and Table \ref{resultTable} for the northern hemisphere fall 
2020 class. As can be seen,  90\%
of students surveyed thought that \AHaH\ was good to excellent in helping understanding of lab 
materials, 92\% thought it was good to excellent in making the lab more 
interesting, and 91\%
thought that it was good to excellent in answering questions they had on the subject 
matter. Finally, throughout the semester students consistently expressed to instructors a preference for 
activities with \AHaH, as well as greater ease of understanding with the tool. Teaching 
Assistants regularly reported notably higher amounts of positive feedback on AHaH-related 
labs as compared to standard lab exercises.

\begin{sidewaystable}
\caption{Questions from survey used in northern hemisphere Fall 2020 AST 113/SES 123 program.}
\label{questionTable}
\begin{tabularx}{\linewidth}{L L}
\toprule
{Question} & {Response Type}\\
\midrule
    I understand that my responses will be used -- without my name and in aggregate -- in a journal article, and I consent to this use. & 
    Acknowledgement \\
    \midrule
    How well did you feel AHaH and related labs aided your understanding of the content covered by these labs? 
    & Multiple Choice on scale of 1 (Poor) to 5 (Excellent)  \\
    \midrule
    How well did you feel AHaH and related labs helped in making labs more interesting? 
    & Multiple Choice on scale of 1 (Poor) to 5 (Excellent)  \\
    \midrule
    How well did you feel AHaH and related labs helped answer your questions about this material? 
    & Multiple Choice on scale of 1 (Poor) to 5 (Excellent) \\
    \midrule
    Other comments? 
    & Free response \\
\bottomrule
\end{tabularx}
\end{sidewaystable}

\begin{sidewaystable}
\caption{Summary of results from northern hemisphere fall 2020 program, with number and percentage
of responses to questions, and summary of comments from students.  Each of the 
two categories above were rated on a scale of 1-5, with 1 being not at all and 5 
being excellently useful in this regard.  We consider here responses of 5 to be 
``Excellent," 3-4 to be ``Good," and 1-2 to be ``Had Issues."}
\label{resultTable}
\begin{tabularx}{\linewidth}{L L L L}
\toprule
Category & Excellent & Good & Had Issues\\
\midrule
    Understanding of Lab Materials & 32 (58\%) & 18 (32\%) & 5 (9\%) \\
    \midrule
    Interesting for Lab Exercise & 31 (56\%) & 20 (36\%) & 4 (7\%) \\
    \midrule
    Answered Questions & 31 (56\%) & 19 (35\%) & 5 (9\%) \\
    \midrule
    Comments
     & Excited to use tool as visualization - many students were interested in the differences between galaxies visually, as well as getting an idea of differences in distances.
     & Interface took getting used to - some students expressed an interest in the tool being redesigned as an app, or with more immediate information pop-ups. 
     & Common difficulties were with installation of the tool. \\
\bottomrule
\end{tabularx}
\end{sidewaystable}

\section{Other Program Uses}
\label{uses}

In addition to in the classroom, \AHaH\ has been and continues to be useful in a
wide range of applications in STEM education. We have conducted several pilot
programs to use this tool in outreach efforts through several events held on and
about ASU campus. As in education, while static images have utility in outreach
efforts with the general public, people young and old are far more likely to
become and stay engaged with dynamic media, such as simulation. As previous
studies have shown (\eg\ \citealt*{Holzinger2008}), simple dynamic media can improve the
acquisition and retention of information. In the case of interactive simulation
like \AHaH, this is likely because the user becomes directly involved in shaping
the course of their experience. In our program, this is exemplified by members of
the public being able to pick out the galaxies they ``fly" towards using \AHaH,
and can learn more details about these. These programs have shown a qualitative increase in
participation by the general public in outreach activities, especially among
young children or young adults. As many public educators will attest, half the battle is often
getting the public to start using to an education opportunity. Thus the use of \AHaH\ in the 
classroom is worth consideration by  teachers, outreach developers, and organizers.

\section{Conclusion}
\label{conclusion}

We believe that our \AHaH\ software provides students and instructors with an
unique ability to interactively visualize many of the effects of a
relativistically expanding Universe, among its other capabilities. The
application should help clarify these concepts, and allow students to develop a
deeper intuitive understanding of the material. Certain cosmological effects --
such as bandpass shifting, {\it k}-correction, surface brightness dimming,
gravitational lensing, and the effects of the magnitude limit and object sizes
on the sample completeness limit -- have been largely omitted due to computational
limitations, but we believe these to be not essential for the understanding of
the included effects. For a discussion of these more technical effects, see \eg,
\citet{Windhorst2018}. In addition, our brief study of the utility of the
program in the classroom meets our expectations of its impact on learning, and
lends support to our recommendation that further virtual tools be developed in
support of online classrooms. We also recommend the use of this tool in other
public education and science outreach efforts for its utility in quickly
engaging the public.

For the convenience of those who wish to see or modify the particular
implementation of the above formulae within the Java software, we have provided
source code with the standard distribution of the tool. It is included in the
src/ directory of ahah.jar, and may be extracted using the java \texttt{jar}
utility or any zlib-compatible de-compressor such as \texttt{unzip}. The tool
may be downloaded from the \AHaH\
website\footnote{\url{http://ahah.asu.edu}}. Further details on \AHaH\ download 
instructions and installation are given in the \AHaH\ user manual, available on the tool website.

\section{Availability of source code and requirements}

\begin{itemize}
    \item Project name: Appreciating Hubble At Hyperspeed
    \item Project home page: \url{http://ahah.asu.edu}
    \item Operating system(s): Microsoft Windows, Mac OS, or *nix OS
    \item Programming language: Java
    \item Other requirements: Java 1.4.3 or higher, 1 GHz processor, 256 MB RAM, Mouse and Keyboard
    \item License: BSD-like.
\end{itemize}

Virtually any modification or redistribution of the application is permitted, with the following caveats:
\begin{itemize}
    \item Any redistribution must retain the original copyright notice and license file, either with the source code or with the documentation in the case of binary distributions.
    \item The names of the copyright holders, contributors, and associated institutions may not be used to endorse or promote any derivative works without prior permission.
\end{itemize}

The application's source code is provided with the standard distribution.

\section{Availability of supporting data and materials}

The data used to develop the teaching tool is available through the sources we cite throughout the paper. As part of our Institutional Review Board agreement, our data may only be published and discussed in aggregate, so we cannot make available our raw survey data.

\section{Declarations}

\subsection{List of abbreviations}
\begin{itemize}
    \item ACS : Advanced Camera for Surveys
    \item \AHaH\ : Appreciating Hubble at Hyperspeed
    \item ASU : Arizona State University
    \item CTIO : Cerro Tololo Inter-American Observatory
    \item GRAPES : Grism-ACS Program for Extragalactic Science
    \item HUDF : Hubble Ultra Deep Field
    \item HST : Hubble Space Telescope
    \item IRB : Institutional Review Board
    \item ISAAC : Infrared Spectrometer And Array Camera
    \item NICMOS : Near Infrared Camera and Multi-Object Spectrometer
    \item VLT : Very Large Telescope
\end{itemize}

\subsection{Ethical Approval}
As noted above, we filed for and received approval with the ASU Institutional Review Board to conduct our survey in the AST 113 and SES 123 Lab classrooms, and their approval, containing all conditions for the study, was filed with ASU.  Students were informed of the new content that would be part of the class, and that there would be a non-graded, optional survey at the end of class concerning the \AHaH\ tool.  All participants in the survey, as indicated in our procedures, gave consent that they were over 18 years of age and that their responses would be reported in aggregate.

\subsection{Consent for publication}

Not applicable, see above.

\subsection{Competing Interests}

The authors declare that they have no competing interests.

\subsection{Funding}

We acknowledge student support from the Arizona State University NASA Space Grant (to LJN and MRM). We acknowledge support from Hubble Space Telescope grants HST-GO-10530.07-A, HST-GO-13779.005-A, HST-EO-10530.25-A and HST-EO-13241.001-A from STScI, which is operated by AURA for NASA under contract NAS 5-26555. RAW acknowledges support from NASA JWST Interdisciplinary Scientist grants NAG-12460, NNX14AN10G and 80NSSC18K0200 from GSFC.

\subsection{Author's Contributions}

LN was the general project lead, developed the survey and other in-classroom components, completed IRB procedures, and wrote the majority of the text on background and program applications.  MM led development of the \AHaH\ code and wrote much of the initial technical paper on the tool's function.  RW was in charge of general project management and oversight, as well as geometric formalism, development of related labs, and assisted with paper modification.  KK, LW, TA, and RW were in charge of various labs and accommodating the \AHaH\ survey.  SC helped with the data preparation for the tool, and assisted in paper development.  ST gave general project assistance.

\section{Acknowledgements}

We thank Dr. Ned Wright for helpful discussion early in the project. We sincerely appreciate comments from both reviewers of this paper, which helped us better anchor our work in the historical context, and better suit the audience of AEJ and current educational research. We also thank the wonderful Teaching Assistants of AST 113 and SES 123: Angelica Berner, Katherine Elder, Ebraheem Farag, Connor Gelb, Jake Hanson, Isabela Huckabee, Darby Kramer, Kelley Liebst, and Kyle Massingill. Much of our education design was based in part on the work by \citet{Hasper2015}.

\del{\section{Authors' information (optional)}

You may choose to use this section to include any relevant information about the author(s) that may aid the reader's interpretation of the article, and understand the standpoint of the author(s). This may include details about the authors' qualifications, current positions they hold at institutions or societies, or any other relevant background information. Please refer to authors using their initials. Note this section should not be used to describe any competing interests.}

\bibliography{paper-refs}

\begin{thebibliography}{}

\bibitem[{Beckwith} et~al., 2006]{Beckwith2006}
{Beckwith}, S. V.~W., {Stiavelli}, M., {Koekemoer}, A.~M., {Caldwell}, J.
  A.~R., {Ferguson}, H.~C., {Hook}, R., {Lucas}, R.~A., {Bergeron}, L.~E.,
  {Corbin}, M., {Jogee}, S., {Panagia}, N., {Robberto}, M., {Royle}, P.,
  {Somerville}, R.~S., and {Sosey}, M. (2006).
\newblock {The Hubble Ultra Deep Field}.
\newblock {\em AJ}, 132(5):1729--1755.

\bibitem[{Bertin} and {Arnouts}, 1996]{Bertin1996}
{Bertin}, E. and {Arnouts}, S. (1996).
\newblock {SExtractor: Software for source extraction.}
\newblock {\em AAPS}, 117:393--404.

\bibitem[{Bertin} et~al., 2002]{Bertin2002}
{Bertin}, E., {Mellier}, Y., {Radovich}, M., {Missonnier}, G., {Didelon}, P.,
  and {Morin}, B. (2002).
\newblock {The TERAPIX Pipeline}.
\newblock In {Bohlender}, D.~A., {Durand}, D., and {Handley}, T.~H., editors,
  {\em Astronomical Data Analysis Software and Systems XI}, volume 281 of {\em
  Astronomical Society of the Pacific Conference Series}, page 228.

\bibitem[{Bolzonella} et~al., 2000]{Bolzonella2000}
{Bolzonella}, M., {Miralles}, J.~M., and {Pell{\'o}}, R. (2000).
\newblock {Photometric redshifts based on standard SED fitting procedures}.
\newblock {\em AAP}, 363:476--492.

\bibitem[{Bouwens} et~al., 2006]{Bouwens2006}
{Bouwens}, R.~J., {Illingworth}, G.~D., {Blakeslee}, J.~P., and {Franx}, M.
  (2006).
\newblock {Galaxies at z \raisebox{-0.5ex}\textasciitilde 6: The UV Luminosity
  Function and Luminosity Density from 506 HUDF, HUDF Parallel ACS Field, and
  GOODS i-Dropouts}.
\newblock {\em APJ}, 653(1):53--85.

\bibitem[Coble et~al., 2018]{Coble2018}
Coble, K., Conlon, M., and Bailey, J.~M. (2018).
\newblock Investigating undergraduate students' ideas about the curvature of
  the universe.
\newblock {\em Phys. Rev. Phys. Educ. Res.}, 14:010144.

\bibitem[{Cohen} et~al., 2006]{Cohen2006}
{Cohen}, S.~H., {Ryan}, R.~E., J., {Straughn}, A.~N., {Hathi}, N.~P.,
  {Windhorst}, R.~A., {Koekemoer}, A.~M., {Pirzkal}, N., {Xu}, C., {Mobasher},
  B., {Malhotra}, S., {Strolger}, L.~G., and {Rhoads}, J.~E. (2006).
\newblock {Clues to Active Galactic Nucleus Growth from Optically Variable
  Objects in the Hubble Ultra Deep Field}.
\newblock {\em APJ}, 639(2):731--739.

\bibitem[Cole et~al., 2018]{Cole2018}
Cole, M., Cohen, C., Wilhelm, J., and Lindell, R. (2018).
\newblock Spatial thinking in astronomy education research.
\newblock {\em Phys. Rev. Phys. Educ. Res.}, 14:010139.

\bibitem[{Dahlen} et~al., 2007]{Dahlen2007}
{Dahlen}, T., {Mobasher}, B., {Dickinson}, M., {Ferguson}, H.~C., {Giavalisco},
  M., {Kretchmer}, C., and {Ravindranath}, S. (2007).
\newblock {Evolution of the Luminosity Function, Star Formation Rate,
  Morphology, and Size of Star-forming Galaxies Selected at Rest-Frame 1500 and
  2800 {\r{A}}}.
\newblock {\em APJ}, 654(1):172--185.

\bibitem[{Ellis}, 2007]{Ellis2007}
{Ellis}, G. F.~R. (2007).
\newblock {On the definition of distance in general relativity: I. M. H.
  Etherington (Philosophical Magazine ser. 7, vol. 15, 761 (1933))}.
\newblock {\em General Relativity and Gravitation}, 39(7):1047--1052.

\bibitem[{Ellis}, 2009]{Ellis2009}
{Ellis}, G. F.~R. (2009).
\newblock {Re-publication of: Relativistic cosmology (1971), In: General
  Relativity and Cosmology, Proc. of the International School of Physics
  “Enrico Fermi,” R.K. Sachs (ed), (New York: Academic Press)}.
\newblock {\em General Relativity and Gravitation}, 41(3):581--660.

\bibitem[{Eriksson} et~al., 2014]{Eriksson2014}
{Eriksson}, U., {Linder}, C., {Airey}, J., and {Redfors}, A. (2014).
\newblock {Introducing the anatomy of disciplinary discernment: an example from
  astronomy}.
\newblock {\em European Journal of Science and Mathematics Education},
  2:167--182.

\bibitem[{Etherington}, 2007]{Etherington2007}
{Etherington}, I.~M.~H. (2007).
\newblock {Re-publication of: LX. On the definition of distance in general
  relativity (1933), Phil. Mag., 15, 761.}
\newblock {\em General Relativity and Gravitation}, 39(7):1055--1067.

\bibitem[{Harrison}, 1993]{Harrison1993}
{Harrison}, E. (1993).
\newblock {The Redshift-Distance and Velocity-Distance Laws}.
\newblock {\em APJ}, 403:28.

\bibitem[Hasper et~al., 2015]{Hasper2015}
Hasper, E., Windhorst, R., Hedgpeth, T., Van~Tuyl, L., Gonzales, A., Martinez,
  B., Yu, H., Farkas, Z., and Baluch, D. (2015).
\newblock {Methods for Creating and Evaluating 3D Tactile Images to Teach STEM
  Courses to the Visually Impaired}.
\newblock {\em J. College Sc. Teaching}, 44.

\bibitem[Hoeling, 2012]{Hoeling2012}
Hoeling, B. (2012).
\newblock Interactive online optics modules for the college physics course.
\newblock {\em American Journal of Physics}, 80:334.

\bibitem[Hogg, 1999]{Hogg1999}
Hogg, D. (1999).
\newblock Distance measures in cosmology.

\bibitem[Holzinger et~al., 2008]{Holzinger2008}
Holzinger, A., Kickmeier-Rust, M., and Albert, D. (2008).
\newblock Model-based gaussian and non-gaussian clustering.
\newblock {\em Educational Technology \& Society}, 11:279--290.

\bibitem[{Kristian} and {Sachs}, 1966]{Kristian1966}
{Kristian}, J. and {Sachs}, R.~K. (1966).
\newblock {Observations in Cosmology}.
\newblock {\em APJ}, 143:379.

\bibitem[Longair, 1995]{Longair1995}
Longair, M. (1995).
\newblock The deep universe.
\newblock {\em Saas-Fee Advanced Course23}, page 317.

\bibitem[Longair, 1998]{Longair1998}
Longair, M. (1998).
\newblock {\em {Galaxy formation}}.
\newblock New York : Springer.

\bibitem[{Lupton} et~al., 2004]{Lupton2004}
{Lupton}, R., {Blanton}, M.~R., {Fekete}, G., {Hogg}, D.~W., {O'Mullane}, W.,
  {Szalay}, A., and {Wherry}, N. (2004).
\newblock {Preparing Red-Green-Blue Images from CCD Data}.
\newblock {\em PASP}, 116(816):133--137.

\bibitem[Peebles, 1993]{Peebles1993}
Peebles, P. J.~E. (1993).
\newblock {\em Principles of physical cosmology}.
\newblock Princeton University Press.

\bibitem[{Pirzkal} et~al., 2017]{Pirzkal2017}
{Pirzkal}, N., {Malhotra}, S., {Ryan}, R.~E., {Rothberg}, B., {Grogin}, N.,
  {Finkelstein}, S.~L., {Koekemoer}, A.~M., {Rhoads}, J., {Larson}, R.~L.,
  {Christensen}, L., {Cimatti}, A., {Ferreras}, I., {Gardner}, J.~P.,
  {Gronwall}, C., {Hathi}, N.~P., {Hibon}, P., {Joshi}, B., {Kuntschner}, H.,
  {Meurer}, G.~R., {O'Connell}, R.~W., {Oestlin}, G., {Pasquali}, A., {Pharo},
  J., {Straughn}, A.~N., {Walsh}, J.~R., {Watson}, D., {Windhorst}, R.~A.,
  {Zakamska}, N.~L., and {Zirm}, A. (2017).
\newblock {FIGS{\textemdash}Faint Infrared Grism Survey: Description and Data
  Reduction}.
\newblock {\em APJ}, 846(1):84.

\bibitem[{Pirzkal} et~al., 2004]{Pirzkal2004}
{Pirzkal}, N., {Xu}, C., {Malhotra}, S., {Rhoads}, J.~E., {Koekemoer}, A.~M.,
  {Moustakas}, L.~A., {Walsh}, J.~R., {Windhorst}, R.~A., {Daddi}, E.,
  {Cimatti}, A., {Ferguson}, H.~C., {Gardner}, J.~P., {Gronwall}, C., {Haiman},
  Z., {K{\"u}mmel}, M., {Panagia}, N., {Pasquali}, A., {Stiavelli}, M., {di
  Serego Alighieri}, S., {Tsvetanov}, Z., {Vernet}, J., and {Yan}, H. (2004).
\newblock {GRAPES, Grism Spectroscopy of the Hubble Ultra Deep Field:
  Description and Data Reduction}.
\newblock {\em APJS}, 154(2):501--508.

\bibitem[{Planck Collaboration} et~al., 2018]{Planck2018}
{Planck Collaboration}, {Aghanim}, N., {Akrami}, Y., {Ashdown}, M., {Aumont},
  J., {Baccigalupi}, C., {Ballardini}, M., {Banday}, A.~J., {Barreiro}, R.~B.,
  {Bartolo}, N., {Basak}, S., {Battye}, R., {Benabed}, K., {Bernard}, J.~P.,
  {Bersanelli}, M., {Bielewicz}, P., {Bond}, J.~R., {Borrill}, J., {Bouchet},
  F.~R., {Burigana}, C., {Calabrese}, E., {Carron}, J., {Chiang}, H.~C.,
  {Comis}, B., {Contreras}, D., {Crill}, B.~P., {Curto}, A., {Cuttaia}, F., {de
  Bernardis}, P., {de Rosa}, A., {de Zotti}, G., {Delabrouille}, J., {Di
  Valentino}, E., {Dickinson}, C., {Diego}, J.~M., {Dor{\'e}}, O., {Ducout},
  A., {Dupac}, X., {Elsner}, F., {En{\ss}lin}, T.~A., {Eriksen}, H.~K.,
  {Falgarone}, E., {Fantaye}, Y., {Finelli}, F., {Forastieri}, F., {Frailis},
  M., {Fraisse}, A.~A., {Franceschi}, E., {Frolov}, A., {Galeotta}, S.,
  {Galli}, S., {Ganga}, K., {Gerbino}, M., {G{\'o}rski}, K.~M., {Gruppuso}, A.,
  {Gudmundsson}, J.~E., {Handley}, W., {Hansen}, F.~K., {Herranz}, D., {Hivon},
  E., {Huang}, Z., {Jaffe}, A.~H., {Keih{\"a}nen}, E., {Keskitalo}, R.,
  {Kiiveri}, K., {Kim}, J., {Kisner}, T.~S., {Krachmalnicoff}, N., {Kunz}, M.,
  {Kurki-Suonio}, H., {Lamarre}, J.~M., {Lasenby}, A., {Lattanzi}, M.,
  {Lawrence}, C.~R., {Le Jeune}, M., {Levrier}, F., {Liguori}, M., {Lilje},
  P.~B., {Lindholm}, V., {L{\'o}pez-Caniego}, M., {Lubin}, P.~M., {Ma}, Y.~Z.,
  {Mac{\'\i}as-P{\'e}rez}, J.~F., {Maggio}, G., {Maino}, D., {Mandolesi}, N.,
  {Mangilli}, A., {Martin}, P.~G., {Mart{\'\i}nez-Gonz{\'a}lez}, E.,
  {Matarrese}, S., {Mauri}, N., {McEwen}, J.~D., {Melchiorri}, A., {Mennella},
  A., {Migliaccio}, M., {Miville-Desch{\^e}nes}, M.~A., {Molinari}, D.,
  {Moneti}, A., {Montier}, L., {Morgante}, G., {Natoli}, P., {Oxborrow}, C.~A.,
  {Pagano}, L., {Paoletti}, D., {Partridge}, B., {Perdereau}, O., {Perotto},
  L., {Pettorino}, V., {Piacentini}, F., {Plaszczynski}, S., {Polastri}, L.,
  {Polenta}, G., {Rachen}, J.~P., {Racine}, B., {Reinecke}, M., {Remazeilles},
  M., {Renzi}, A., {Rocha}, G., {Roudier}, G., {Ruiz-Granados}, B., {Sandri},
  M., {Savelainen}, M., {Scott}, D., {Sirignano}, C., {Sirri}, G., {Spencer},
  L.~D., {Stanco}, L., {Sunyaev}, R., {Tauber}, J.~A., {Tavagnacco}, D.,
  {Tenti}, M., {Toffolatti}, L., {Tomasi}, M., {Tristram}, M., {Trombetti}, T.,
  {Valiviita}, J., {Van Tent}, F., {Vielva}, P., {Villa}, F., {Vittorio}, N.,
  {Wandelt}, B.~D., {Wehus}, I.~K., {Zacchei}, A., and {Zonca}, A. (2018).
\newblock {Planck intermediate results. LIII. Detection of velocity dispersion
  from the kinetic Sunyaev-Zeldovich effect}.
\newblock {\em AAP}, 617:A48.

\bibitem[{Retzlaff} et~al., 2010]{Retzlaff2010}
{Retzlaff}, J., {Rosati}, P., {Dickinson}, M., {Vandame}, B., {Rit{\'e}}, C.,
  {Nonino}, M., {Cesarsky}, C., and {GOODS Team} (2010).
\newblock {The Great Observatories Origins Deep Survey. VLT/ISAAC near-infrared
  imaging of the GOODS-South field}.
\newblock {\em AAP}, 511:A50.

\bibitem[{Ribeiro}, 2005]{Ribeiro2005}
{Ribeiro}, M.~B. (2005).
\newblock Cosmological distances and fractal statistics of galaxy distribution.
\newblock {\em A\&A}, 429(1):65--74.

\bibitem[{Ryan} et~al., 2007]{Ryan2007}
{Ryan}, R.~E., J., {Hathi}, N.~P., {Cohen}, S.~H., {Malhotra}, S., {Rhoads},
  J., {Windhorst}, R.~A., {Budav{\'a}ri}, T., {Pirzkal}, N., {Xu}, C.,
  {Panagia}, N., {Moustakas}, L.~A., {di Serego Alighieri}, S., and {Yan}, H.
  (2007).
\newblock {The Galaxy Luminosity Function at z
  \raisebox{-0.5ex}\textasciitilde= 1 in the HUDF: Probing the Dwarf
  Population}.
\newblock {\em APJ}, 668(2):839--845.

\bibitem[Ryden, 2017]{Ryden2017}
Ryden, B. (2017).
\newblock {\em {Introduction to Cosmology, 2nd Edition}}.
\newblock Cambridge University Press, Cambridge, United Kingdom.

\bibitem[Sadaghiani, 2011]{Sadaghiani2011}
Sadaghiani, H.~R. (2011).
\newblock Using multimedia learning modules in a hybrid-online course in
  electricity and magnetism.
\newblock {\em Phys. Rev. ST Phys. Educ. Res.}, 7:010102.

\bibitem[{Thompson} et~al., 2005]{Thompson2005}
{Thompson}, R.~I., {Illingworth}, G., {Bouwens}, R., {Dickinson}, M.,
  {Eisenstein}, D., {Fan}, X., {Franx}, M., {Riess}, A., {Rieke}, M.~J.,
  {Schneider}, G., {Stobie}, E., {Toft}, S., and {van Dokkum}, P. (2005).
\newblock {The Near-Infrared Camera and Multi-Object Spectrometer Ultra Deep
  Field: Observations, Data Reduction, and Galaxy Photometry}.
\newblock {\em AJ}, 130(1):1--12.

\bibitem[Weinberg, 1972]{Weinberg1972}
Weinberg, S. (1972).
\newblock {\em Gravitation and cosmology}.
\newblock Wiley.

\bibitem[{Windhorst} et~al., 2011]{Windhorst2011}
{Windhorst}, R.~A., {Cohen}, S.~H., {Hathi}, N.~P., {McCarthy}, P.~J., {Ryan},
  Russell~E., J., {Yan}, H., {Baldry}, I.~K., {Driver}, S.~P., {Frogel}, J.~A.,
  {Hill}, D.~T., {Kelvin}, L.~S., {Koekemoer}, A.~M., {Mechtley}, M.,
  {O'Connell}, R.~W., {Robotham}, A. S.~G., {Rutkowski}, M.~J., {Seibert}, M.,
  {Straughn}, A.~N., {Tuffs}, R.~J., {Balick}, B., {Bond}, H.~E., {Bushouse},
  H., {Calzetti}, D., {Crockett}, M., {Disney}, M.~J., {Dopita}, M.~A., {Hall},
  D. N.~B., {Holtzman}, J.~A., {Kaviraj}, S., {Kimble}, R.~A., {MacKenty},
  J.~W., {Mutchler}, M., {Paresce}, F., {Saha}, A., {Silk}, J.~I., {Trauger},
  J.~T., {Walker}, A.~R., {Whitmore}, B.~C., and {Young}, E.~T. (2011).
\newblock {The Hubble Space Telescope Wide Field Camera 3 Early Release Science
  Data: Panchromatic Faint Object Counts for 0.2-2 {\ensuremath{\mu}}m
  Wavelength}.
\newblock {\em APJS}, 193(2):27.

\bibitem[{Windhorst} et~al., 2018]{Windhorst2018}
{Windhorst}, R.~A., {Timmes}, F.~X., {Wyithe}, J. S.~B., {Alpaslan}, M.,
  {Andrews}, S.~K., {Coe}, D., {Diego}, J.~M., {Dijkstra}, M., {Driver}, S.~P.,
  {Kelly}, P.~L., and {Kim}, D. (2018).
\newblock {On the Observability of Individual Population III Stars and Their
  Stellar-mass Black Hole Accretion Disks through Cluster Caustic Transits}.
\newblock {\em APJS}, 234(2):41.

\bibitem[{Wright}, 2006]{Wright2006}
{Wright}, E.~L. (2006).
\newblock {A Cosmology Calculator for the World Wide Web}.
\newblock {\em PASP}, 118(850):1711--1715.

\bibitem[{Yan} and {Windhorst}, 2004]{YanWindhorst2004}
{Yan}, H. and {Windhorst}, R.~A. (2004).
\newblock {Candidates of z \raisebox{-0.5ex}\textasciitilde= 5.5-7 Galaxies in
  the Hubble Space Telescope Ultra Deep Field}.
\newblock {\em APJL}, 612(2):L93--L96.

\end{thebibliography}

\section{Appendix B.  Derivation of Equations} \label{derivations}
Here we include our derivations of the math used in \AHaH\ at a level 
appropriate for an intermediate undergraduate cosmology class\footnote{See, \eg, \url{http://windhorst322.asu.edu}}.
\subsection{Comoving Radial Distance} \label{cmr dist}

To begin, we need the comoving radial distance, $D_R$, from the Earth to an
object at redshift $z$, derived from the Robertson-Walker metric, as discussed
previously, \eg, \citet[Ch.~7][]{Longair1998}, Eqs.~5.33 and 6.13 of
\citet{Ryden2017}, and Eq.~6 of \citet{Wright2006}. We express this as the
integral:

\begin{equation}\label{D_R}
 D_R(z) = \int_{t}^{t_0} \frac{c \cdot dt}{a} = 
 \int_{\frac{1}{1+z}}^{1} \frac {c \cdot da}{a \dot a} = 
 \frac{c}{H_0} \int_{0}^{z} \frac{dz}{(1+z) \dot a},
\end{equation}

\n where the scale factor $a=1/(1+z)$. The derivative of $a$ with respect to time,
$\dot a$, is given by the expression:

\begin{equation}\label{a_dot}
 \dot a = (\Omega_M/a + \Omega_R/a^2 + \Omega_\Lambda \cdot a^2 +
\Omega_K)^{1/2},
\end{equation}

\n where $\Omega_{M}$, $\Omega_{R}$, $\Omega_{\Lambda}$, and $\Omega_{K}$ are
energy density parameters, corresponding to the fractions of the Universe's total
average energy density that are attributable to matter (M), radiation (R), dark
energy ($\Lambda$), and the curvature of the spatial geometry (K), respectively.
Note that it is assumed these are the only relevant contributions to the total
energy density $\Omega_{Tot}$. That is, we assume that $\Omega_{Tot}$ = $\Omega_M
+ \Omega_\Lambda + \Omega_R + \Omega_K$. A spatially flat Universe would have 
$\Omega_{Tot} \equiv 1$ with $\Omega_K$ = 0. The default \cite{Planck2018}
parameters used are: $H_{o}$ = 68 km/sec/Mpc, $\Omega_{M}$ = 0.32,
$\Omega_{\Lambda}$ = 0.68, $\Omega_{R}$ = 9$\times$10$^{-5}$ with $\Omega_K$ = 0.

We evaluate this integral in steps of 0.05 in $z$ from $z=0$ to $z=20$ to
create a look-up table, interpolating linearly to find the value for any
arbitrary redshift in between these discrete steps. This is because we must
make the calculation frequently and for many objects, so computing the integral
manually every time would be computationally prohibitive. The resulting error
in this method is generally small enough that it translates to less than one
pixel's difference even on high-resolution displays, so this error can safely
be ignored for the purposes of the application. We evaluate the integral using
the simple midpoint method, which may not be the optimal solution, but was
simple to implement and adequately efficient on any home computer. As with the
linear interpolation, higher accuracy numerical integration would result in
less than one pixel's difference when displayed.

\subsection{Angular Size Distance} \label{angular dist}

To develop the angular size distance, $D_A$, we first need to develop a 
generalized form of the comoving radial distance $D_R$, to express the distance measure to an object at 
redshift $z_j$ as measured by an observer at redshift $z_i$. This distance is 
given by the formula:

\begin{equation}\label{D_T}
D_R(z_i, z_j) = \left\{ \begin{array}{cc}
\Re_i \sin(r_i/{\Re_i}) &\textrm{ if } \Omega_K < 0 \\
r_i &\textrm{ if } \Omega_K = 0 \\
\Re_i \sinh(r_i/{\Re_i}) &\textrm{ if } \Omega_K > 0 \\
\end{array}\right.,
\end{equation}

\n where $\Re'$ is the radius of curvature of the spatial geometry at redshift
$z_i$, and $r_i$ is the value of the comoving coordinate distance at the
same redshift \citep{Longair1998, Wright2006}. These correspond to the cases
where the spatial geometry of the Universe is spherically curved, flat, and
hyperbolically curved, respectively. Recalling that both $r'_{ij}$ and $\Re'$
scale as $1/(1+z_i)$, and that $\Re=(c/H_0)/\sqrt{|\Omega_K|}$, we next define an
intermediate quantity $U$, representing the argument of $\sin$ and $\sinh$ in 
\eqn{D_T} above:

\begin{equation}\label{Quantity U}
U = r_i/\Re_i = r_0/\Re_0 = (H_0/c)\sqrt{|\Omega_K|}r_0
\end{equation}

We note that since $U$ now depends only upon the cosmology selected by the user 
and the object's redshift, we may calculate $U$ once per object and re-use it, 
thus saving CPU time. Using this quantity, we may now rewrite $D_R$ as:

\begin{equation}\label{New D_T}
D_R(z_i, z_j) = \frac{\delta(U)}{1+z_i}r_0
\end{equation}

Here, $\delta(U)$ is simply some function of $U$. By substituting $U$ into 
\eqn{D_T} above, we get the following expression for $\delta(U)$:

\begin{equation}\label{Delta}
 \delta(U) = \left\{ \begin{array}{cc}
 \frac {\sin(U)} {U}  & \textrm{ if } \Omega_K < 0 \\
 1				      & \textrm{ if } \Omega_K = 0 \\
 \frac {\sinh(U)} {U} & \textrm{ if } \Omega_K > 0 \\
\end{array}\right.
\end{equation}

\n Note that $\delta(U)$ expressly depends upon $r_0$. The case where $\Omega_K=0$ 
comes from the limit of both $\sin(U)/U$ and $\sinh(U)/U$ as $\Omega_K\to0$ -- 
one may observe that in this case \eqn{New D_T} simplifies to $r_0/(1+z_i)$, which 
is precisely $r_i$ as in \eqn{D_T}, since as observers we start our \AHaH\ journey 
at $z_i$ = 0. When we expand with \AHaH\ into the HUDF galaxy database as it is 
sorted in redshift, the observer's redshift can take any value 0\cle $z_i$\cle 20, 
although the current set of HUDF postage stamp images does not have any galaxies 
with z\cge 6. We intend to expand this with a future version of \AHaH\ based on the 
full data set summarized in \citet{Windhorst2011}. 

Thus, using our \eqn{New D_T} and the equation relating the angular size distance 
and the distance measure as developed by \citet[Eq.~7.50]{Longair1998}, the angular 
size distance from redshift $z_i$ to $z_j$ is given by:

\begin{equation}\label{D_A}
D_A(z_i, z_j) = D_R(z_i, z_j)\frac{1+z_i}{1+z_j} = \frac{\delta(U)}{1+z_j}r_0
\end{equation}

\subsection{Comoving Coordinate System} \label{coordinates}

Now that we have developed formulae for $D_R$ and $D_A$, we can consider the best
way to create a coordinate system for the Java application. The data we start
with are the redshift $z_j$ of object $j$ (with which we can calculate $D_R$) and
four angles: the object's angular size (from the height and width of its image)
and the angular separation between the object and the $x$ and $y$ axes, which we
define as lines going through the center of the original image. These angles are
calculated by taking the corresponding size in pixels and multiplying by the
scale in arcsec/pixel of the original HST image\footnote{The HUDF mosaics used
have been drizzled to a scale of 0\arcspt 03 per pixel}.

We would like to use this information to create a coordinate system with the
original telescope position at the origin. In a Euclidean space this would
present no problem, but we have already remarked that the \textit{observed}
angles are {\it not} the same in an expanding Universe as they would be in a
Euclidean space. Further, it would be desirable for the Euclidean coordinate
distance to correspond to the comoving radial distance, as this would make
calculations significantly simpler. We can accomplish this, but when we create
coordinates for each object as such, we need to ``correct'' the observed
angles. That is, we want a ``Euclidean angular size'' associated with a certain
observed angular size. We will call this $\theta_E$. An object's angular size
is related to its physical transverse diameter, $d$, by the following equation which derives from \eqn{D_A}:

\begin{equation}\label{Theta-Diameter}
 d = \theta D_A = \theta \frac{\delta(U)}{1+z_j}r_0 = \theta_E 
	 \frac{1}{1+z_i}r_0
\end{equation}

\n Note that in the Euclidean case we must contract $r_0$ by a factor of
$1/(1+z_i)$ to get the comoving distance from $z_i$ to $z_j$ as measured by the
observer at $z_i$ ($r_i$ in \eqn{D_T} above). This is because the proper
spatial separation in the current epoch has been stretched by the Universe's
expansion, so as seen by an observer at redshift $z_i$ it must be scaled
appropriately. Hence, the equivalent Euclidean distance between any two points 
is $D_E = r_0/(1+z_i)$, in which case we get the Euclidean small-angle 
approximation back: $d = \theta_E D_E$.

Thus canceling $r_0$, we get the following expression for $\theta_E$ from \eqn{Theta-Diameter}:

\begin{equation}\label{Theta_E}
 \theta_E = \theta\delta(U)\frac{1+z_i}{1+z_j}
\end{equation}

\n In our initial data $z_i$ is simply zero, so that we create coordinates ($X$,
$Y$, $Z$) for an object at redshift $z$ like:

\begin{equation}\label{X Definition}
 X = \sin\left(\frac{\delta(U)\theta_X}{1+z}\right)
	 \cos\left(\frac{\delta(U)\theta_Y}{1+z}\right)D_R(0,z),
\end{equation}


\n and similarly for $Y$ and $Z$. We have thus developed a coordinate system of
$X$, $Y$, and $Z$ in comoving Mpc with the original telescope position at the origin.

\subsection{Simulating Observations From Vantage Points Other Than \texorpdfstring{$z=0$}{z=0}} 
\label{simulated obs}

\n Now, when we ``move'' the Hubble camera virtually to higher redshifts, we do
so by moving to some new ($X_c$, $Y_c$, $Z_c$) value in the coordinate space.
(Note that \AHaH\ does not only ``virtually violate the laws of physics'' by
moving the observer into the HUDF images at $\sim$500$\times$10$^{12}$ times the
speed of light, but it also has no problem ``violating the arrow of time'' by
allowing the user to move back and forth in redshift through the sorted HUDF
image data cube). By construction, the distance measure here is just the
Euclidean coordinate distance:

\begin{equation}\label{Euclidean Distance}
 D_E = ((X-X_c)^2+(Y-Y_c)^2+(Z-Z_c)^2)^{1/2}
\end{equation}

\n Now to determine where to display an object after we have ``moved'' the camera,
we use the distance calculated with \eqn{Euclidean Distance} and the Euclidean
angular size. Using the redshift of the object of interest, $z_o$, and the
camera's user-defined redshift, $z_c$, we rearrange \eqn{Theta_E} to get:

\begin{equation}\label{New Angle}
 \theta = \theta_E \frac{1+z_o}{\delta_{ij}(1+z_c)}
\end{equation}

\n Here $\delta_{ij}$ follows from \eqn{D_T} and \eqn{Delta} for an object at redshift $z_j$ as observed from redshift $z_i$. In this case, $\theta_E$ is a quantity that we must calculate from our
coordinates in the usual Euclidean way.

For an object's angular size it is even simpler than for its $(X,Y,Z)$
position, since we do not have to manually calculate $\theta_E$. We know that in
the Euclidean case:

\begin{equation}\label{Angular Size Theta-Diameter Redshift 0}
 d = \theta_0 D_R = \theta_E D_E,
\end{equation}

\n where $\theta_0$ is the Euclidean angular size from redshift zero, and $D_E$ is
the coordinate distance from the camera to the object from \eqn{Euclidean
Distance}. We then solve for $\theta_E$ in \eqn{Angular Size Theta-Diameter
Redshift 0}, $\theta_E=\theta_0D_R/D_E$, and substitute back into \eqn{New Angle} to obtain an expression for the
desired angular size $\theta$ for any object at redshift $z_o$ as observed from
$z_c$:

\begin{equation}\label{Angular Size Theta}
 \theta = \theta_0 \left( \frac{D_R}{D_E} \right) 
 \frac{1+z_o}{\delta_{ij}(1+z_c)}
\end{equation}

\section{Appendix B. Specialized Data Preparation} \label{prep}
\label{data prep}

To develop the images used in \AHaH\ we first created a custom-balanced RGB version of the HUDF image\footnote{\label{hudf}See
\url{http://imgsrc.hubblesite.org/hu/db/2004/07/images/a/formats/full_jpg.jpg} 
for the full-resolution 60 Mb HUDF ACS image in the BViz filters, and 
\url{https://www.asu.edu/clas/hst/www/aas2014/HUDF14-pan-UVrendered.jpg} for
the deepest 13-filter panchromatic 849-orbit Hubble image.}. While the image
provided in the original press releases$^{\ref{hudf}}$ would have been adequate, it has the
undesirable characteristic that very bright areas, such as bulges in large
spiral galaxies, appear burned-out and lack fine detail. The HUDF data used was
taken to roughly equal depths in the $B$-, $V$-, $i'$-, and $z'$-band filters,
which have central wavelengths of $\sim$4320 (Blue), 5920 (Visual), 7690
(Red), and 9030 \AA\ (near-Infrared), respectively \citep{Beckwith2006}, so we
created a three-channel color image by first combining the $B$- and $V$-bands,
applying weights based on the sky signal-to-noise ratio \footnote{We applied a
weight of 0.765 in $V$ and 0.235 in $B$ to balance the image depths. For details, see \citet{Windhorst2011}}. We then
used the algorithm developed by \citet{Lupton2004} to create the combined RGB
image, with the combined $B\!+\!V$-bands as the blue channel, the $i'$-band as
the green channel, and the $z'$-band as the red channel\footnote{The channels
were first scaled proportional to the data zero points -- Red: 716.474, Green:
345.462, Blue: 254.449, see \url{https://hst-docs.stsci.edu/acsihb}.}. Besides
showing more detail in bright areas, this method has the added benefit that an
object with a specified astronomical color has a unique color in the composite
RGB image. A comparison of the original STScI color images and our prepared
images is shown in \fig{figcomparison}. The full HUDF image using this color
preparation technique is also available as an interactive map online\footnote{
\url{http://ahah.asu.edu/clickonHUDF/index.html}}.  One can see examples of galaxies processed in this way in \fig{figthreegalaxies}.

The galaxies represented in the \AHaH\ application were $i'$-band (8000 \AA)  
selected using the \sex\ algorithm \citep{Bertin1996, Bertin2002} with a detection threshold of 1.8$\times$
the rms noise level (1.8$\sigma$) above the local sky. The $i'$-band (at $z\!\approx\!6$) dropouts of
\citet{YanWindhorst2004} were added by hand. We then created color JPEG
``stamp'' images for each individual object, using the \sex-generated
segmentation map to mask as black any pixels outside the detected source. These
``stamps'' were then converted pixel-for-pixel to PNG images, which employ a
lossless compression algorithm -- no image quality was thus lost. We then
developed a transparency map based on each pixel's brightness, which was saved
into the PNG alpha channel\footnote{The alpha channel of a PNG contains the ``transparency" of each pixel, which is more straightforwardly the opacity of the pixel.  For more, see the PNG specification page from w3: \url{https://www.w3.org/TR/PNG-DataRep.html}.}. The resulting images can thus be displayed as {\it
semi-transparent}, allowing objects in the distance to show through the dim
regions of objects in the foreground, as is also possible in the real Universe.

Photometric redshifts for the galaxies were measured with the \texttt{HyperZ}
package \citep{Bolzonella2000}, using a combination of the original HST-ACS
four-band ($BVi'z'$) data from the HUDF, along with $Y$-, $J$- and $H$-band near-IR data
from HST NICMOS (Near Infrared Camera and Multi-Object Spectrometer) \citep{Thompson2005}. 
We have supplemented the photometric redshifts with
spectro-photometric redshifts measured by \citet{Ryan2007}, which incorporate
the aforementioned $BVi'z'JH$ data, as well as grism spectra from GRAPES
\citep{Pirzkal2004, Pirzkal2017}, the $U$-band observations from CTIO (Cerro Tololo Inter-American Observatory) Mosaic II \citep{Dahlen2007},
and $K_s$-band data from VLT-ISAAC (Very Large Telescope -- Infrared Spectrometer And Array Camera, e.g. \citet{Retzlaff2010}). For a summary of all these data and the
quality of the spectro-photometric redshifts, see \citet{Ryan2007}. When
available, we have chosen to use the more reliable spectro-photometric redshifts.

\end{document}